\documentclass[aps,prd,twocolumn]{revtex4}
\usepackage{graphicx}% Include figure files
\usepackage{dcolumn}% Align table columns on decimal point
\def\beq{\begin{equation}}
\def\eeq{\end{equation}}
\def\bea{\begin{eqnarray}}
\def\eea{\end{eqnarray}}

\def\v{\varphi}
\def\H{{\cal H}}
\def\sgn{{\rm sgn}}
\def\bwt{\begin{widetext}}
\def\ewt{\end{widetext}}

\begin{document}

\title{Cuscuton: A Causal Field Theory with an Infinite Speed of Sound}
\author{Niayesh Afshordi}\email{nafshordi@cfa.harvard.edu}\affiliation{Institute for
  Theory and Computation,
Harvard-Smithsonian Center for Astrophysics, MS-51, 60 Garden
Street, Cambridge, MA 02138}
\author{Daniel J.H. Chung}\email{Danielchung@wisc.edu}
\affiliation{Department of Physics, University of Wisconsin,
  Madison, WI 53706}
\author{Ghazal Geshnizjani}\email{ghazal@physics.wisc.edu}
\affiliation{Department of Physics, University of Wisconsin,
  Madison, WI 53706}
\date{\today}
\preprint{hep-th/yymmnnn}
\begin{abstract}
We introduce a model of scalar field dark energy, {\it Cuscuton},
which can be realized as the incompressible (or infinite speed of
sound) limit of a scalar field theory with a non-canonical kinetic
term (or k-essence). Even though perturbations of {\it Cuscuton}
propagate superluminally, we show that they have a locally
degenerate phase space volume (or zero entropy), implying that they
cannot carry any microscopic information, and thus the theory is
causal. Even coupling to ordinary scalar fields cannot lead to
superluminal signal propagation. Furthermore, we show that the
family of constant field hypersurfaces are the family of Constant
Mean Curvature (CMC) hypersurfaces, which are the analogs of soap
films (or soap bubbles) in a Euclidian space. This enables us to
find the most general solution in 1+1 dimensions, whose properties
motivate conjectures for global degeneracy of the phase space in
higher dimensions. Finally, we show that the {\it Cuscuton} action
can model the continuum limit of the evolution of a field with
discrete degrees of freedom and argue why it is protected against
quantum corrections at low energies. While this paper mainly focuses
on interesting features of {\it Cuscuton} in a Minkowski spacetime,
a companion paper \cite{companion} examines cosmology with {\it
Cuscuton} dark energy.

%has been attributed to a mysterious dark energy or a large scale
%modification of the theory of gravity. While a cosmological constant
%has been so far consistent with all cosmological observations,
%various models of dynamical/evolving dark energy have been put forth

\end{abstract}
%\pacs{13.15.+g, 64., 64.30.+t, 64.70., 64.70.Fx, 98.80., 98.80.Cq}
\maketitle

\section{Introduction}

There has been much recent interest in field theories with
non-canonical kinetic terms. Many of these theories are inspired by
purely phenomenological motivations. K-essence
\cite{Armendariz-Picon:2000dh,Chiba:1999ka} and k-inflation
 \cite{Armendariz-Picon:1999rj} have been designed in order to solve
the cosmological coincidence and inflationary fine tuning problems,
while Bekenstein's theory of gravity \cite{Bekenstein:2004ne}
attempts to accommodate apparent deviations from Newtonian gravity
on galactic scales, within a relativistically covariant framework.

More theoretical motivations have led to theories such as ghost
condensate \cite{Arkani-Hamed:2003uy}, which is considered as an
analog of Higgs boson for general relativity, or variations of
Born-Infeld action that describe non-perturbative objects in string
theory \cite{Silverstein:2003hf,Sen:2002nu}.

Here we present a new class of actions with a non-canonical kinetic
term which is characterized by the Hamiltonian symplectic structure
of the theory degenerating in a cosmologically interesting
homogeneous limit. In other words, in the limit that the field
degree of freedom becomes homogeneous in the locally freely falling
frame (where the metric is locally a Minkowski metric), the equation
of motion does not have any second order time derivatives and the
field becomes a non-dynamical auxiliary field, which merely follows
the dynamics of the fields that it couples to. Thus we call this
field {\it Cuscuton} (pronounced {\tt k\"{a}s-k\"{u}-t\"{a}n}),
after the Latin name for the parasitic plant of dodder, Cuscuta
\footnote{The word Cuscuta is most likely of Aramaic or Hebrew
origin, which has entered Latin through Arabic or Greek
\cite{cusc}.}.

Nonetheless, the {\it Cuscuton} action may a priori appear to have
dynamical degrees of freedom because there is a non-degenerate
conjugate momentum degree of freedom if the field configuration has
a non-vanishing spatial gradient.  It is this feature which justifies
exploration of field theoretic features of this class of theories
presented here.

For concreteness, we will focus on scalar field actions of the
k-essence \cite{Armendariz-Picon:2000dh,Armendariz-Picon:2000ah}
form\begin{equation} S_{\varphi}=\int
d^{4}x\sqrt{-g}\left[\frac{1}{2}F(X,\varphi)-V(\varphi)\right],\label{eq:simpformstart}\end{equation}
where $X\equiv\partial_{\mu}\varphi\partial^{\mu}\varphi$ with a
particular class of choice for $F$ such as to satisfy our
degeneration of symplectic structure feature that will be defined in
the next section.

There has been some controversy in the literature regarding the
causality of dynamical k-essence models with $c_s > 1$. While the
original literature on k-essence (e.g.
\cite{Armendariz-Picon:2000dh,Armendariz-Picon:2000ah}), as well as
some follow-up studies \cite{2002astro.ph.12302D,Bruneton:2006gf},
argue that superluminal modes of k-essence cannot carry information
on closed loops, and thus do not break causality, others
\cite{Adams:2006sv,Bonvin:2006vc} take $c_s>1$ at its face value,
arguing that it cannot be realized as the IR limit of an inherently
causal field theory. Superluminal scalar field models are also
claimed to lead to signal propagation out of the horizon of a black
hole \cite{Babichev:2006vx}, as well as large tensor-to-scalar
ratios for inflationary perturbations \cite{Mukhanov:2005bu}.

In Section\ref{causality}, we show that although at face value {\it
Cuscuton} seems to possess dynamics, which allows superluminal
signal propagation, it actually contains no local dynamical degree
of freedom for $X>0$. What that means is that the theory even with a
kinetic term acts like a pure constraint system, modifying the
dynamics of whatever it couples to. In that sense, this is a
k-essence theory which behaves like modified gravity, since if it
couples to gravity and other fields, it merely provides a means of
changing those dynamical sectors through constraints without adding
any new local degrees of freedom of its own. Moreover, the
superluminal modes cannot carry any information due to this
degeneration of dynamical degrees of freedom \footnote{However, note
that although the corner that smoothly connects to the phase space
of degenerating symplectic structure contains no local dynamics,
this is not the case in the region where the variation of
$S_{\varphi}$ with respect to $X$ becomes singular. We leave this
for a future investigation since that part of the theory is not
fixed by the definition of the class of theories that we are
proposing.}.

Section \ref{CMC discussion} demonstrates that constant field
hypersurfaces have constant mean curvature, making them Minkowski
space analogs of soap bubbles/films in a Euclidian space. This leads
to exact solutions of theory in 1+1 dimensions in Section
\ref{exact}, which enables us to make general conjectures about
uniqueness and generic singularity of the solutions. We go on to
briefly consider the coupling of {\it Cuscuton} to gravity (in the
context of homogeneous cosmology), as well as ordinary scalar
fields, in Section \ref{coupled}. The latter, in particular, shows
that propagating degrees of freedom remain subluminal, even after
coupling to {\it Cuscuton}.

In Section \ref{instanton} we discuss a possible physical model for
{\it Cuscuton action}, and argue why it may be protected against
quantum corrections at low energies. Finally, Section~\ref{conclude}
summarizes our results and concludes the paper. A companion paper
\cite{companion} examines cosmology with a {\it Cuscuton} dark
energy fluid.

%Throughout this paper we use the reduced Planck's constant
%$M_{p}=(8\pi G_{N})^{-1/2}\approx2\times10^{18}$GeV.

\section{Defining {\it Cuscuton} Action}\label{define_cus}

Here, we define the class of models of interest for this paper.
Readers interested in only concrete examples should skip to Eq.
(\ref{eq:simplerform}). Consider the action of the form Eq.
(\ref{eq:simpformstart}) in a Minkowski spacetime and choose $F$
such that in the homogeneous limit of the field the kinetic term
becomes a total derivative for $\dot{\varphi}\neq0$ (and thus would
drop out of the equation of motion):\begin{eqnarray}
F(X,\varphi(x))\rightarrow F(\dot{\varphi}^{2},\varphi(t)) & = & \frac{d}{dt}J(\varphi,\dot{\varphi},...)\nonumber\\
 & = & \dot{\varphi}\frac{\partial J}{\partial\varphi}+\ddot{\varphi}\frac{\partial J}{\partial\dot{\varphi}}+....\end{eqnarray}
Since $F$ does not contain any $\ddot{\varphi}$ or higher derivative
functions, we conclude $J=J(\varphi)$ and \begin{equation}
F(\dot{\varphi}^{2},\varphi)=\sqrt{\dot{\varphi}^{2}}\frac{\partial
J(\varphi)}{\partial\varphi}\label{eq:fparticular},\end{equation}
where we have absorbed the sign of $\dot{\varphi}$ into
$\frac{\partial J(\varphi)}{\partial\varphi}$ and $F$ is well
defined with respect to field variation as long as the sign of
$\dot{\v}$ does not change. %except at the point $\dot{\varphi}=0$.
Hence, the action in the homogeneous limit is simply\begin{equation}
S_{\varphi}^{homog}= -\int d^{4}xV(\varphi),\end{equation} (up to
boundary terms) which when coupled to another field, for example
$\chi$, gives the total action\begin{equation} S_{\textrm{example
}\chi}=\int
d^{4}x[\mathcal{L}_{\chi}(\chi,\varphi(t))-V(\varphi(t))]\end{equation}
giving rise to a constraint equation\begin{equation} -\frac{\partial
V}{\partial\varphi}+\frac{\partial\mathcal{L}_{\chi}(\chi,\varphi)}{\partial\varphi}=0.\label{eq:exampcoup}\end{equation}
This modifies the dynamics of $\chi$ depending on the choice of
function $V(\varphi)$. As long as $\chi$ has dynamics to make
$\dot{\varphi}\neq0$, the field $\varphi$ acts like a non-dynamical
auxiliary field.

Just because the homogeneous limit with $\dot{\varphi}\neq0$ is
non-dynamical does not mean a priori that $\varphi$ does not have
any dynamics, especially since the classical equation of motion in
Minkowski space takes the form (for $X>0$ with
$F(X,\varphi)=\sqrt{X}\frac{\partial J(\varphi)}{\partial\varphi}$
coming from Eq. (\ref{eq:fparticular}))\begin{equation}
\frac{1}{\sqrt{-g}}\partial_{\mu}\left[\frac{\sqrt{-g}}{2}\frac{\partial^{\mu}\varphi}{\sqrt{X}}\frac{\partial
J}{\partial\varphi}\right]-{\sqrt{X}\over
2}\frac{\partial^{2}J}{\partial\varphi^{2}}+V'(\varphi)=0,\label{eq:minkowskieom}\end{equation}
which clearly has a second time derivative as long as
$\partial_{i}\varphi\neq0$. More formally, the existence of dynamics
can be studied in the Hamiltonian formalism \begin{equation} \{ {\sf
H}(\varphi,\Pi),\varphi(x)\}=\partial_{0}\varphi(x)\label{eq:hamilton1},\end{equation}
\begin{equation}
\{ {\sf
H}(\varphi,\Pi),\Pi(x)\}=\partial_{0}\Pi(x)\label{eq:hamilton2},\end{equation}
\begin{equation}
\{\Pi(t,\vec{x}),\varphi(t,\vec{y})\}=\delta^{(3)}(\vec{x}-\vec{y}),\label{eq:commutator}\end{equation}
where ${\sf H}\equiv\int d^{3}x\mathcal{H}$ is the Hamiltonian,
$\{,\}$ is the functional Poisson brackets, and $\Pi$ is the
conjugate momentum to $\varphi$. This set of equations preserves the
classical phase space whose volume element can be defined
symplectically in the form $D\Pi\wedge D\varphi.$ In the case of Eq.
(\ref{eq:fparticular}), the Hamiltonian density for the theory is:
\beq \H = \sgn\left(\frac{\partial J}{\partial\v}
\right)|\nabla\v|\sqrt{\Pi^2-\frac{1}{4}\left(\frac{\partial
J}{\partial\v}\right)^2}+V(\v), \label{hamilton}\eeq while the
canonical momentum $\Pi$ takes the form \begin{equation}
\Pi=\frac{1}{2}\frac{\partial^{0}\varphi}{\sqrt{X}}\frac{\partial
J}{\partial\varphi},\end{equation} which should satisfy Eqs.
(\ref{eq:hamilton1}-\ref{eq:commutator}), and have a normal phase
space element $D\Pi\wedge D\varphi$. However, in the limit that
$\partial_{i}\varphi\rightarrow0$, $\Pi$ becomes only dependent on
$\varphi$, as long as $\dot{\varphi}(t)$ does not cross zero,
i.e.\begin{equation}
\Pi(t)=\frac{1}{2}\textrm{sgn}(\dot{\varphi})\frac{\partial
J(\varphi(t))}{\partial\varphi}\end{equation}
 and the symplectic element $D\Pi\wedge D\varphi$ collapses (since
$D\varphi\wedge D\varphi=0$). That is simply a signature of
$\varphi$ becoming non-dynamical in the homogeneous limit -- i.e.
has no phase space -- except at the singular point
$\dot{\varphi}=0$.

To summarize, we define the {\it Cuscuton} action as the class of
actions in which $\varphi$ becomes non-dynamical in the homogeneous
limit. %of at least one Lorentz frame in the regime
%$\dot{\varphi}\neq0$.
The most general such action, with a single real scalar field, and a
covariant kinetic term with no more than first order gradients (as
assumed in Eq. (\ref{eq:simpformstart})) is
\begin{equation} S_{\varphi}=\int
d^{4}x\sqrt{-g}\left[\frac{1}{2}\frac{\partial
J(\varphi)}{\partial\varphi}\sqrt{|g^{\mu\nu}\partial_{\mu}\varphi\partial_{\nu}\varphi|}-V(\varphi)\right].\label{eq:simpleform}\end{equation}
%where we have promoted the metric to a generally covariant metric in
%anticipation of coupling the action to gravity. Whenever we consider
%specific examples, we will work with a simpler version of the action
%with
It is easy to see that, as long as $\frac{\partial J}{\partial \v}
>0$, the field $\v$ can be re-defined  to set
\begin{equation} \frac{\partial
J(\varphi)}{\partial\varphi} = 2\mu^{2} = {\rm const.,
}\label{eq:simpcoeff}\end{equation} although the choice for the
value of $\mu^2$ is arbitrary. Thus, for the rest of the paper, we
will use the following form for the {\it Cuscuton} action:\beq
S_{\varphi}=\int
d^{4}x\sqrt{-g}\left[\mu^2\sqrt{|g^{\mu\nu}\partial_{\mu}\varphi\partial_{\nu}\varphi|}-V(\varphi)\right].\label{eq:simplerform}\end{equation}

%(where $\mu^{2}$ is a constant) to simplify the relevant
%illustrations.

Note that we have inserted the absolute value in the radicand of
Eq.~(\ref{eq:simplerform}) to make the action well defined when
$X\equiv
g^{\mu\nu}\partial_{\mu}\varphi\partial_{\nu}\varphi<0$. However, when
$X<0$, $\partial_{\mu}\varphi$ is space-like, implying that we cannot
use a Lorentz boost to reach a locally homogeneous (and thus
non-dynamical) limit. Hence, the absolute value inside Eq.~(\ref{eq:simplerform}) is an ansatz that does not follow from our
general definition of {\it Cuscuton} models, and has been chosen
merely for simplicity. Finally, note that even with an absolute value
in Eq.~(\ref{eq:simplerform}), the variation of the action is
ill-defined at $X=0$. In such cases, the definition of the equation of
motion from the direct consideration of the action is necessary.  For
example, the path integral of the form
\begin{equation} Z=\int D\varphi e^{iS_{\varphi}}\end{equation} is
likely to be well defined even though the variation of $\delta
S_{\varphi}/\delta\varphi$ does not exist at $X=0$. We will not deal
with this issue further in this paper. %although we will speculate on
%implications for effective field theory in a latter section.

\section{Why Is {\it Cuscuton} Causal?}\label{causality}

At the linearized level, most dynamical theories are characterized
by second order partial differential equations whose characteristic
curves delimit the support for Green's functions which propagate
Cauchy data. If the characteristic curves allow propagation of
information outside of the lightcone (defined by the local Lorentz
group), one may worry that the theory is acausal, leading to an
ill-defined initial value formulation.  With this reasoning, the
condition for well defined causal Cauchy data problem of linearized
second order partial differential equations coming from actions of
the form given in Eq. (\ref{eq:simpformstart}) are given by
Aharonov, Komar, and Susskind \cite{Aharonov:1969vu}. The conditions
for causal structure and energy positivity to be preserved
are\begin{equation} F'(X)>0\label{eq:firstcauscond}\end{equation}
\begin{equation}
F''(X)\geq0\label{eq:secondcauscond}\end{equation}
 where $'$ denotes derivative with respect to $X$, while
the stability of solutions with respect to small changes in Cauchy
data requires \begin{equation}
2XF''/F'>-1.\label{eq:cauchydata}\end{equation}

The latter condition is intimately related to the definition of
speed of sound, $c_s$, in scalar field theories with non-canonical
kinetic terms (or k-essence)
\cite{Armendariz-Picon:2000dh,Armendariz-Picon:2000ah}:
\begin{eqnarray} c_{s}^{2} &
= & \frac{1}{1+2XF''/F'}.\label{soundspeed}\end{eqnarray} Thus, the
condition for stability of solutions (Eq. (\ref{eq:cauchydata})) is
equivalent to
 $c^2_s>0$, while the causality conditions (\ref{eq:firstcauscond}-\ref{eq:secondcauscond}) ensure $c_s\leq 1$.

 With $F$
chosen to be Eq. (\ref{eq:fparticular}), Eq.
(\ref{eq:firstcauscond}) gives\begin{equation}
\frac{1}{2}\frac{1}{\sqrt{X}}\frac{\partial
J}{\partial\varphi}>0\label{eq:cond1}\end{equation}
 while Eq. (\ref{eq:secondcauscond}) gives\begin{equation}
\frac{-1}{4}\frac{1}{X^{3/2}}\frac{\partial
J}{\partial\varphi}\geq0.\label{eq:cond2}\end{equation} It is clear
that both of these conditions cannot be satisfied. Hence, we would
be naturally concerned that this class of theories are acausal,
especially since substituting Eq.~(\ref{eq:fparticular}) in
Eq.~(\ref{soundspeed}) yields $c_s = \infty$, or an infinite speed
of sound! However, as we will now argue, upon closer inspection, the
theory of Eq.~(\ref{eq:simpformstart}) has no problems with
causality from a local signal propagation point of view.

The first argument comes from a linearized analysis. The simplified
equation of motion Eq.~(\ref{eq:minkowskieom}) has a second order
differential operator that can be rewritten as\begin{equation}
(\eta^{\mu\nu}-\frac{\partial^{\mu}\varphi\partial^{\nu}\varphi}{X})\partial_{\mu}\partial_{\nu}\varphi,\end{equation}
which means that the characteristic curves for the linearized
equation (accounting only for the highest derivative operator) are
governed by the effective metric\begin{equation}
\tilde{g}_{\mu\nu}\equiv\eta_{\mu\nu}-\frac{\partial_{\mu}\varphi_{0}\partial_{\nu}\varphi_{0}}{\partial_{\alpha}\varphi_{0}\partial^{\alpha}\varphi_{0}},\label{field_metric}\end{equation}
where the linearization is about a background field configuration
$\varphi_{0}$: i.e. $\varphi=\varphi_{0}+\delta\varphi$. Writing out
a coordinate dependent expression for $\tilde{g}_{\mu\nu}$, we
see\bwt\begin{equation}
\tilde{g}_{\mu\nu}=\frac{1}{\dot{\varphi}^{2}-(\vec{\nabla}\varphi)^{2}}\left(\begin{array}{cc}
-(\vec{\nabla}\varphi)^{2} & -\partial_{0}\varphi_{0}\partial_{i}\varphi\\
-\partial_{0}\varphi_{0}\partial_{i}\varphi &
-[\dot{\varphi}^{2}-(\vec{\nabla}\varphi)^{2}]\delta_{ij}-\partial_{i}\varphi\partial_{j}\varphi\end{array}\right),\end{equation}\ewt
where the Latin indices denote the spatial components and run over
1-3. All diagonal components of the effective metric are manifestly
negative and by direct computation, one sees that the determinant of
this matrix is $0$. Hence, the theory is manifestly Euclidean at the
linearized level (requiring only Dirichlet initial data on a
three-surface and not Cauchy data). Consequently, as far as the
linear theory is concerned, there is no dynamics which means that
$\delta\varphi$ merely satisfies a constraint equation. %and is not
%characterizing the state of the system. Hence
Even though the effective metric seems to allow characteristic
curves to carry information about $\delta\varphi$ outside of the
lightcone, there is no information carried by $\delta\varphi$
independently of the fields to which $\delta\varphi$ couples. On the
other hand, the fields to which $\varphi$ couples, say $\chi$, can
change their causal properties due to the constraint equation.
However, this is a model dependent problem requiring the analysis of
the propagator for $\chi$ subject to the solution of the constraint
equation. For example, we show in a companion paper \cite{companion}
that metric perturbations coupled to a {\it Cuscuton} field evolve
causally, as the scalar curvature does not change on super-horizon
scales.

The second argument, which is a generalization of the argument
above, is that despite the generally non-vanishing $\Pi$, because of
the underlying Lorentz symmetry, one can always go to a frame in
which $D\Pi\wedge D\v  =0$ locally, and hence, there is no local
dynamics to this system.

To understand what we mean by the collapse of the phase space
structure, let us consider a simple toy model particle dynamical
system consisting of the Lagrangian\begin{equation}
L(q_{i},\dot{q}_{i})=\sqrt{\dot{q}_{1}^{2}+\dot{q}_{2}^{2}}-V(q_{1},q_{2}).\end{equation}
 Here, one can heuristically think of $\dot{q}_{1}$ as $\partial_{0}\varphi(t,\vec{x}_{1})$
and $\dot{q}_{2}$ as $\partial_{0}\varphi(t,\vec{x}_{2})$ with
$\vec{x}_{1}\neq\vec{x}_{2}$. In that sense, when
$\dot{q}_{2}\neq\dot{q}_{1}$, we have $\partial_{i}\varphi\neq0$.
Now, the conjugate momentum to $q_{i}$ is \begin{equation}
p_{i}=\frac{\dot{q}_{i}}{\sqrt{\dot{q}_{1}^{2}+\dot{q}_{2}^{2}}}.\end{equation}
Hence one sees an analogous degenerate structure in which when
$\dot{q}_{1}=\dot{q}_{2}$, the phase space structure
$\prod_{i}dq_{i}\wedge dp_{i}$ collapses. However, this is secretly
a theory with limited dynamics. To see this, consider making a
change of integration variables of the phase space from
$(q_{1},p_{1},q_{2},p_{2})$ to
$(q_{1},\dot{q}_{1},q_{2},\dot{q}_{2})$. The Jacobian for the
transformation of the measure is\begin{equation}
\det\left[\frac{\partial(q_{1},p_{1},q_{2},p_{2})}{\partial(q_{1},\dot{q}_{1},q_{2},\dot{q}_{2})}\right]=\textrm{det}\left(\begin{array}{cccc}
1 & 0 & 0 & 0\\
0 & \frac{\dot{q}_{2}^{2}}{X^{3/2}} & 0 & \frac{-\dot{q}_{1}\dot{q}_{2}}{X^{3/2}}\\
0 & 0 & 1 & 0\\
0 & \frac{-\dot{q}_{1}\dot{q}_{2}}{X^{3/2}} & 0 &
\frac{\dot{q}_{1}^{2}}{X^{3/2}}\end{array}\right)=0,\end{equation}
where $X\equiv\dot{q}_{1}^{2}+\dot{q}_{2}^{2}$. Hence, one sees that
this is manifestly a theory with degenerate Hamiltonian dynamics,
where in fact the solutions only occupy a 3D hypersurface of the
full 4D phase space of $(q_i,p_i)$.

It is straight forward to generalize the change of phase space
measure in the field theory case. The Jacobian relating
$D\varphi\wedge D\Pi$ to $D\varphi\wedge D\dot{\varphi}$ is\bwt
\begin{equation} \det\frac{\partial(\varphi({\bf x})\Pi({\bf
x'}))}{\partial(\varphi({\bf y})\dot{\varphi}({\bf
y'}))}=\det\left(\begin{array}{cc}
\delta^{(3)}({\bf x -y })& 0\\
 \frac{\Pi}{X}\nabla\v({\bf x'})\cdot\nabla\delta^{(3)}({\bf x'-y})& \frac{-\mu^2|\nabla\varphi({\bf x'})|^{2}}{X^{3/2}}\delta^{(3)}({\bf
 x'-y'})\end{array}\right).\end{equation}\ewt
Now, because Lorentz symmetry allows us to locally rotate
$\nabla\varphi({\bf x'})=0$ for any vector $\partial^{\mu}\varphi$
such that $\partial^{\mu}\varphi\partial_{\mu}\varphi>0$, the
symplectic structure of the phase space collapses without any local
dynamical degrees of freedom. Note that this is in many ways just a
corollary to the perturbative analysis of the theory.

The degeneration of the local phase space volume also implies that
local perturbations do not carry any microscopic information, or
equivalently {\it Cuscuton} fluid has zero entropy.

Here, we should point out that, although lack of internal dynamics
prevents any transfer of information through the {\it Cuscuton}
field, even superluminal propagation does not necessarily lead to a
breakdown of causality \cite{Bruneton:2006gf}. However, co-existence
of interacting k-essence fields, which allow superluminal signal
propagations at different rest frames (and thus different
chronologies) can generically yield the propagation of signals on
closed time-like curves, which does imply a breakdown of causality
\footnote{``time-like" is here defined with respect to the
background field metric(s), Eq. (\ref{field_metric}), rather than
the space-time metric.}. In contrast, in Sec. \ref{klein} we will
show that even (stable) coupling of {\it Cuscuton} to an ordinary
field with propagating degrees of freedom cannot lead to
superluminal signal propagation.

Let us close this section by commenting on the causal properties for
$X<0$. Note that in this case, plugging from  Eq.
(\ref{eq:simpleform}), the conditions of Eqs.
(\ref{eq:firstcauscond}) and (\ref{eq:secondcauscond}) become
\begin{equation} F'=\frac{1}{2}\frac{\partial
J}{\partial\varphi}\frac{\textrm{sgn}X}{\sqrt{|X|}}>0,\end{equation}
\begin{equation}
F''=-\frac{1}{4}\frac{\partial
J}{\partial\varphi}\frac{1}{|X|^{3/2}}\geq0,\end{equation} which
merely require $\partial J/\partial\varphi<0$ for causality. Hence,
with an absolute value ansatz for $\sqrt{|X|}$ part of the
Lagrangian, it would be prudent to impose the $\partial
J/\partial\varphi<0$ condition to keep the theory causal in the
regime $X<0$ (see the Appendix for an explicit demonstration). Note
that this will also keep the kinetic energy density positive
definite in the $X<0$ regime.

To summarize, we argued that since the {\it Cuscuton} field theory
does not seem to have any internal dynamics, despite its apparently
acausal structure, it cannot be used to send information. In the
regime of $X<0$, even though not directly relevant to our analysis,
we can recover internal dynamics, a causal structure, and a positive
definite energy, but only if $\partial J/\partial\varphi<0$.

\section{Constant Mean Curvature (CMC) surfaces: Analogy to Soap Films}\label{CMC discussion}
In this section, we will show that hypersurfaces of constant $\v$
have constant mean curvature, and thus are Minkowski space analogs
of soap films (or soap bubbles) in the Euclidian space.

For the simplified system of Eq. (\ref{eq:simplerform}), the
equation of motion is
\begin{equation}
\frac{1}{\sqrt{-g}}\partial_{\mu}\left[\sqrt{-g}\frac{\partial^{\mu}\varphi}{\sqrt{|\partial^{\mu}\varphi\partial_{\mu}\varphi|}}\right]
+\frac{V'(\varphi)}{\mu^{2}}=0.\label{field}\end{equation}

 To get
more insight into the geometrical nature of solutions, we should
point out that
\begin{equation}
 u^{\mu}\equiv{\partial^{\mu}\varphi \over
 \sqrt{\partial^{\nu}\varphi\partial_{\nu}\varphi}},\end{equation}
are normal unit vectors to constant $\varphi$ hypersurfaces. On the
other hand, the trace of the extrinsic curvature tensor
$K_{\mu\nu}$, or mean curvature of a surfaces, is defined as
\begin{equation}
K\equiv K^\mu_\mu =\nabla_\mu u^\mu,\label{ex_curv}
\end{equation}
which in combination with Eq. (\ref{field}) simply implies that the
mean curvature on constant $\varphi$ hypersurfaces is only a
function of $\varphi$ and hence constant:
\begin{equation}
K(\varphi)=-\frac{V'(\varphi)}{\mu^{2}}.\label{kphi}
\end{equation}

Constant Mean Curvature (CMC) surfaces have been subject of
extensive study in both in mathematics and physics, for their
important and useful features. For example, we should note that a
constant mean curvature surface in Euclidean space can be viewed as
a surface where the exterior pressure and the surface tension forces
are balanced. This can be seen explicitly by looking at the {\it
Cuscuton} action itself (Eq. \ref{eq:simplerform}), which can be
rewritten as:
\begin{eqnarray}
S_{\varphi}&=&\int
d^{4}x\sqrt{-g}[\mu^2\sqrt{|g^{\mu\nu}\partial_{\mu}\varphi\partial_{\nu}\varphi|}-V(\varphi)]\nonumber\\
&=&\int
d^{4}x\sqrt{-g}[\mu^2{|g^{\mu\nu}\partial_{\mu}\varphi\partial_{\nu}\varphi|\over
\sqrt{|g^{\mu\nu}\partial_{\mu}\varphi\partial_{\nu}\varphi|}}-V(\varphi)]\nonumber\\
&=&\int d^{4}x\sqrt{-g}[\mu^2|u^\mu\partial_{\mu}\varphi|-V(\varphi)]\nonumber\\
&=&\mu^2\int_\varphi d\varphi~\Sigma(\varphi)-\int d^{4}x
\sqrt{-g}V(\varphi)\label{sigma}
\end{eqnarray}
where $\Sigma(\varphi)$ is the area of constant $\varphi$ $2+1$
hypersurface in $3+1$ spacetime.

For this reason, CMC surfaces in a Euclidian space can be thought of
as soap bubbles or films (depending on if they have boundaries), as
their configuration is also determined by a similar balance between
surface and volume terms in their energy. In fact, it is easy to
show that our action exactly reproduces such solutions in the case
of $\dot{\varphi}=0$ and $X<0$ and perturbations around
$\dot{\varphi}=0$ or constant $\varphi$ are analogous to propagating
waves on a bubble surface (see Appendix \ref{bubble}).

\section{Exact Solutions, Uniqueness and Singularities}\label{exact}

In general, the question of existence and uniqueness of CMC surfaces
(which constitute the classical solutions to the {\it Cuscuton}
field equation) for a given boundary condition, is of significant
subtlety, and subject of ongoing investigation \cite{MR2013507}.

Nevertheless, we can gain significant insight by studying the
features of this problem in $1+1$ dimensions, where field equation
(\ref{field}) can be exactly solved. In this case, CMC surfaces of
curvature $K$ are in general hyperbolae of the form: \beq
(t-t_0)^2-(x-x_0)^2 = K^{-2}, \eeq where $x_0$ and $t_0$ are
constants.  Note that the hyperbolae degenerate into space-like
lines in the $K\rightarrow 0$ limit. Therefore, using Eq.
(\ref{kphi}), the general solution to the {\it Cuscuton} field
equation is given by: \beq
\left[t-t_0(\v)\right]^2-\left[x-x_0(\v)\right]^2 =
\frac{\mu^4}{V'^2(\v)} \label{exact1+1}\eeq where
${t_0(\varphi),x_0(\varphi)}$ can be multivalued functions of
$\varphi$.

We can now argue that only a discrete set of possible solutions exist
after imposing just the Dirichlet conditions and that general Cauchy
data typically overconstrains the partial differential equation,
resulting in no solution.
%As a result,
%First, note that a given initial condition $\varphi(t=0, x)=\v_0(x)$
%fixes the curvature of all the constant $\varphi$ hyperbolae crossing
%$t=0$.
%Moreover, if $\v_0(x)$ generically takes any given
%value at more than one point, since a CMC hyperbola is completely
%fixed (up to a sign) if it is anchored at two points, there is only
%a finite number of solutions (i.e. family of CMC hyperbolae), that
%would satisfy a given Dirichlet initial condition (and
%$\sgn(\dot{\v})$).
Suppose $\varphi_0(x) \equiv \varphi(t=0,x)$ corresponds to the
initial Dirichlet data.  Suppose there exists a set $S \equiv
\{x_i\}$ which satisfies $\varphi_0(x_i) = v$ where $v$ is a
particular fixed field value occurring in the Dirichlet data. For
any solution $\varphi(t,x)$ to the equation of motion consistent
with the initial Dirichlet data at $t=0$, Eq.~(\ref{exact1+1})
describes the set of constant $\varphi$ hyperbolae slices which {\em
must} pass through {\em all} the points in $\{(t=0,x_i) | x_i \in S
\}$.\footnote{Because of the multivaludedness of $t_0(\varphi)$ and
$x_0(\varphi)$ in Eq.~(\ref{exact1+1}), the map allows multiple
hyperbolas to thread the points in the set $S$.} Since the curvature
$K$ is fixed for each of the hyperbolae (because $v$ is fixed), the
sets of possible hyperbolae form a discrete set.  (Clearly, the set
need not be discrete if the curvatures of the hyperbolae can be
adjusted.)
%On the other hand, an initial condition
%$\v_0(x)$ that does not take any value more than once, cannot
%generically satisfy the solution (\ref{exact1+1}), as any CMC
%hyperbola that crosses the $t=0$ axis, does it twice.
%On the other hand, for a given CMC hyperbola, the tangent of angle
%it crosses the $x$ axis must be $\delta t/\delta x$ or $\alpha
%\equiv\partial_x \varphi/\partial_t\varphi$.
This is remarkable because the set of possible solutions to naively
``hyperbolic'' partial differential equation is a discrete set rather
than a continuous set even when only Dirichlet conditions are imposed.

A corollary to this result is that a Cauchy initial condition, which
fixes both $\v_0$ and $\dot{\v}_0$ (although locally consistent)
typically overconstrains the system globally, and thus is inconsistent
for $K(\varphi)\neq 0$ except for discrete exceptions. Of course, this
is in contrast with regular field theories which require Cauchy
initial conditions to fix their future evolution, and is reminiscent
of local phase space degeneration of {\it Cuscuton}, discussed above
(see \ref{causality}), since a discrete set of points form a set of
measure zero for a continuous measure.

Another corollary of the exact solution (\ref{exact1+1}) is that
%,
%since characteristic hyperbolae are set independently for different
%values of $\v$ at $t=0$, there is nothing preventing them from
since the hyperbolae that thread the points in the set $S$
generically intersect in finite future or past, the normal
derivatives at the intersections are generically not well defined.
Even when a single constant $v$ hyperbolae do not intersect (for
example in the case of there being one hyperbola), for different
values of $\varphi_0(x)$, the hyperbolae will have different
curvatures and thus can intersect, this time making the field value
ill-defined at the intersection.  In these cases, singularities or
discontinuities will generically develop in the solutions within a
finite time.

The development of the field discontinuities is in nature similar to
development of shocks in fluid mechanics, which suggests that they can
be traced using appropriate jump conditions. In our case, the jump
condition is simply a generalization of Eq. (\ref{kphi}): \beq -\mu^2
K_{\rm dis} = \frac{\Delta V}{\Delta \v}, \eeq where $\Delta V$ and
$\Delta \v$ are changes in potential and field values respectively,
across the discontinuity, while $K_{\rm dis}$ is the mean extrinsic
curvature of the discontinuous hypersurface.

We will close this section by commenting on the extension of these
results to $3+1$ dimensions. It is easy to construct a family of
exact solutions by considering CMC hyperbolids of the form: \beq
\left[t-t_0(\v)\right]^2-\left|{\bf x}-{\bf x_0}(\v)\right|^2 =
\frac{9\mu^4}{V'^2(\v)}. \label{exact3+1}\eeq However, this is not
the most general solution to the field equation, as it only
accommodates spherical surfaces of constant $\v$ in 3-space.
Therefore, if the surfaces of constant $\v$ are not spherical in the
field initial conditions, a more general solution should be sought.

Nevertheless, the statement that given Dirichlet initial/boundary
conditions (plus $\sgn(\dot{\v})$) only admit a discrete set of
solutions is still a reasonable conjecture. This conjecture could be
more motivated through the analogy with soap films/bubbles in the
Euclidian space (see \ref{CMC discussion}). For the case of soap
films/bubbles, and assuming a fixed pressure difference, a given
boundary condition only admits a discrete set of solutions
\cite{MR2013507}, which suggests that the same may be true in
Minkowski space.
%Adequacy of Dirichlet initial conditions, and
Generic singularity of the solutions naturally follow from this
conjecture in a similar way to the 1+1 dimensional case.

One can gain another perspective on the solutions to the {\it
Cuscuton} equation of motion Eq.~(\ref{field}) in Minkowski space by
rewriting Eq.~(\ref{exact3+1}) with $x_0^\mu$ set to a constant.  One
can identify $\varphi(x)=f(\Delta x^\mu \Delta x_\mu)$ (where $\Delta
x^\mu \equiv x^\mu - x_0^\mu$) being the solution to the equation of
motion with the potential
\begin{equation}
V(\varphi)= 3 \mu^2 \int \frac{d\varphi}{\sqrt{f^{-1}(\varphi)}}
\end{equation}
with any suitable choice of a single variable function $f(z)$.  For
example, with the choice $f(z)=M e^{-\lambda z^2}$, the potential is
\begin{equation}
V(\varphi)= 3 \mu^2 M \lambda^{1/4} \Gamma(\frac{3}{4}, -\ln \frac{\varphi}{M})
\end{equation}
where $\Gamma(a,b)$ is the incomplete gamma function, and the solution
is
\begin{equation}
\varphi= M e^{-\lambda (\Delta x_\mu  \Delta x^\mu)^2}.
\end{equation}
Notice that this solution is not singular; consistent with our
discussion of singularities, because it does not have intersecting
hyperbolae for two different field values nor does it have two
different hyperbolae characterizing the same field value.

Another example of non-singular solution is in the case of
$V(\varphi)$ being a constant.  In that case, the solution to the
equation of motion is any smooth function
\begin{equation}
\varphi(x)=F(k\cdot x)
\end{equation}
for any constant one-form $k_\mu$.  Again, this solution can be
non-singular because the constant $\varphi$ surfaces are parallel
planes.

To summarize this section, we have
%put forth theorems about the adequacy of
demonstrated that Cauchy boundary conditions generically overconstrain
the equation of motion of {\it Cuscuton} field theory in 1+1
dimensions.  We have also demonstrated in the same theory that the
classical solutions generically have singularities (or
discontinuities) except in some special cases. While we can prove
these theorems in 1+1 dimensions, they remain conjectures in higher
dimensional space-time.

\section{Coupled {\it Cuscuton}}\label{coupled}
In this section, we provide two examples of how coupling to {\it
Cuscuton} can modify the dynamics of physical systems. We first show
how homogenous, or Friedmann-Robertson-Walker (FRW) cosmology is
modified by the presence of {\it Cuscuton} field, with only minimal
coupling to gravity. Then we consider how Klein-Gordon dispersion
relation of perturbations in an ordinary scalar field is modified
through {\it Cuscuton} coupling, and show that propagating degrees
of freedom remain causal.
\subsection{Homogeneous (FRW) {\it Cuscuton} Cosmology}
A comprehensive study of homogeneous cosmology, as well as linear
perturbations and observational constraints can be found in our
companion paper \cite{companion}. Here, for completeness, we provide
a brief treatment of homogeneous (FRW) {\it Cuscuton} cosmology.

The mean extrinsic curvature of comoving hypersurfaces has a
particularly simple interpretation in a cosmological context. Eq.~(\ref{ex_curv}) is also the definition of the expansion rate, or
three times the Hubble constant, $H = \dot{a}/a$, where $a$ is the
scale factor in FRW cosmology. Therefore, Eq. (\ref{kphi}) takes the
form: \beq 3H~\sgn(\dot{\v}) = -\frac{V'(\v)}{\mu^2}.
\label{frw_field}\eeq

Moreover, in Section \ref{define_cus} (Eq.~(\ref{hamilton})), we saw
that the Hamiltonian (or energy) density of {\it Cuscuton}
approaches $V(\v)$ in the homogeneous limit. Therefore, the
Friedmann equation will take the form: \beq H^2 + \frac{\cal K}{a^2}
= \frac{\rho_m + V(\v)}{3M^2_p}, \label{friedmann}\eeq where
$\rho_m$ is the mean matter density of the Universe, $M_p = (8\pi
G)^{-1/2}$ is the reduced Planck mass, and ${\cal K}$ is the
(constant) spatial curvature of the Universe. The {\it Cuscuton}
field, $\v$, can be simply eliminated from Eqs. (\ref{frw_field})
and (\ref{friedmann}), to yield a modified Friedmann equation: \beq
H^2 + \frac{\cal K}{a^2} = \frac{\rho_m +
V\left[V'^{-1}(3\mu^2H)\right]}{3M^2_p}, \label{mod_fried}\eeq where
$V'^{-1}$ is the inverse function of $V'(\v)$ and, without loss of
generality, $\dot{\v} <0$ is assumed. Eq. (\ref{mod_fried})
illustrates the auxiliary nature of {\it Cuscuton}, and that it
simply modifies the dependence of the Hubble expansion rate on
matter density and spatial curvature of the Universe.

A simple example is the quadratic {\it Cuscuton} potential \beq
V(\v) = \frac{1}{2} m^2\v^2, \eeq which yields the modified
Friedmann equation: \beq H^2 +\frac{\cal
K}{\left(1-\frac{3\mu^4}{2m^2M^2_p}\right)a^2} =
\frac{\rho_m}{3\left(M^2_p - \frac{3\mu^4}{2m^2}\right)}, \eeq
implying that quadratic {\it Cuscuton} renormalizes the spatial
curvature and Planck mass to ${\cal K'}$ and $M'_p$, where: \bea
M'^2_p = M^2_p - \frac{3\mu^4}{2m^2},\\ {\cal K'} = \frac{\cal
K}{1-\frac{3\mu^4}{2m^2M^2_p}}. \eea This is a manifestation of why
{\it Cuscuton} can be considered to be a theory of modified gravity
\cite{companion}.

Due to its infinite speed of sound, {\it Cuscuton} does not cluster
on sub-horizon scales, implying the Planck mass approaches its
``fundamental'' value in the UV limit. Therefore, a signature for the
quadratic {\it Cuscuton} model will be a running (or mismatch) of
the Planck mass from the (cosmological) IR to UV regime. Other
cosmological features and observational constraints on the quadratic
{\it Cuscuton} are discussed in \cite{companion}.

\subsection{Coupling to an ordinary scalar field}\label{klein}

Let us study an ordinary scalar field, $\psi$, that is coupled to
{\it Cuscuton} through its potential $V(\v, \psi)$: \beq {\cal L}
(\v,\psi) = \frac{1}{2}\partial_{\alpha}\psi\partial^{\alpha}\psi +
\mu^2\sqrt{|\partial_{\alpha}\v\partial^{\alpha}\v|} - V(\v,\psi).
\eeq

We now study the dispersion relation for linear perturbations
$\delta\psi({\bf x},t)$ and $\delta\v({\bf x},t)$, around a
homogeneous background $\psi(t)$ and $\v(t)$, in the short
wavelength (or WKB) approximation. After decomposing perturbations
into plane wave solutions, or their Fourier components,
$\delta\v_{\omega,{\bf k}}$ and $\delta\psi_{\omega,{\bf k}}$,
%\beq \delta\psi({\bf x},t), \delta\v({\bf x},t) \propto
%\exp\left[i({\bf k\cdot x} -\omega t)\right]. \eeq
the linearized {\it Cuscuton} field equation (\ref{field}) takes the
following form \footnote{Here, we have assumed that the form of the
potential allows for $V_{,\v}=0$ while $\dot{\v}\neq0$.}: \beq
\delta\v_{\omega,{\bf k}} =
-\left(\frac{V_{,\v\psi}}{k^2+V_{,\v\v}}\right)\delta\psi_{\omega,{\bf
k}}, \eeq where, without loss of generality, we have locally
redefined $\v$ to have $|\dot{\v}| = \mu^2$, for the background
$\v(t)$. This can be plugged into the linearized Klein-Gordon
equation for $\psi$: \beq (\omega^2 - k^2) \delta\psi_{\omega,{\bf
k}} = V_{,\psi\psi} \delta\psi_{\omega,{\bf
k}}+V_{,\v\psi}\delta\v_{\omega,{\bf k}}, \eeq to give the modified
dispersion relation for $\delta\psi_{\omega,k}$ perturbations. After
simple manipulations, the dispersion relation takes the form: \beq
\omega^2 =
\frac{k^4+(V_{,\v\v}+V_{,\psi\psi})k^2+V_{,\v\v}V_{,\psi\psi}-V^2_{,\v\psi}}{k^2+V_{,\v\v}}.\label{omega2}
\eeq Notice that again, $\v$ acts as an auxiliary field, and is
dropped out of the field equation, without introducing any
additional degree of freedom.

Assuming a positive determinate Hessian matrix, $V_{,ab}$, for the
second derivatives of the potential $V(\v,\psi)$, we require that
both trace and determinant of $V_{,ab}$ are positive: \bea
V_{,\v\v}+V_{,\psi\psi} >0,\label{trace} \\
V_{,\v\v}V_{,\psi\psi}-V^2_{,\v\psi}
>0, \label{determinant}\eea which ensures that there will be no tachyonic solutions
($\omega^2>0$, i.e. no instability).

To demonstrate causal propagation of {\it dressed} $\delta \psi$
perturbations, we can calculate the group velocity in the
short-wavelength limit: \beq v_g \equiv \frac{d\omega}{dk} = 1 -
\frac{V_{,\psi\psi}}{2k^2} + O\left(V^2_{,\psi\v}\over k^4\right) <
1. \eeq Therefore, given that we require $V_{,\psi\psi} > 0$ from
Eqs. (\ref{trace}-\ref{determinant}), {\it signal propagation will
always be subluminal/causal}, in the regime of validity of plane
wave (WKB) approximation.

\section{ {\it Cuscuton} as an effective action and quantum corrections}
\label{instanton}

\begin{figure*}
\includegraphics[width=0.9\linewidth]{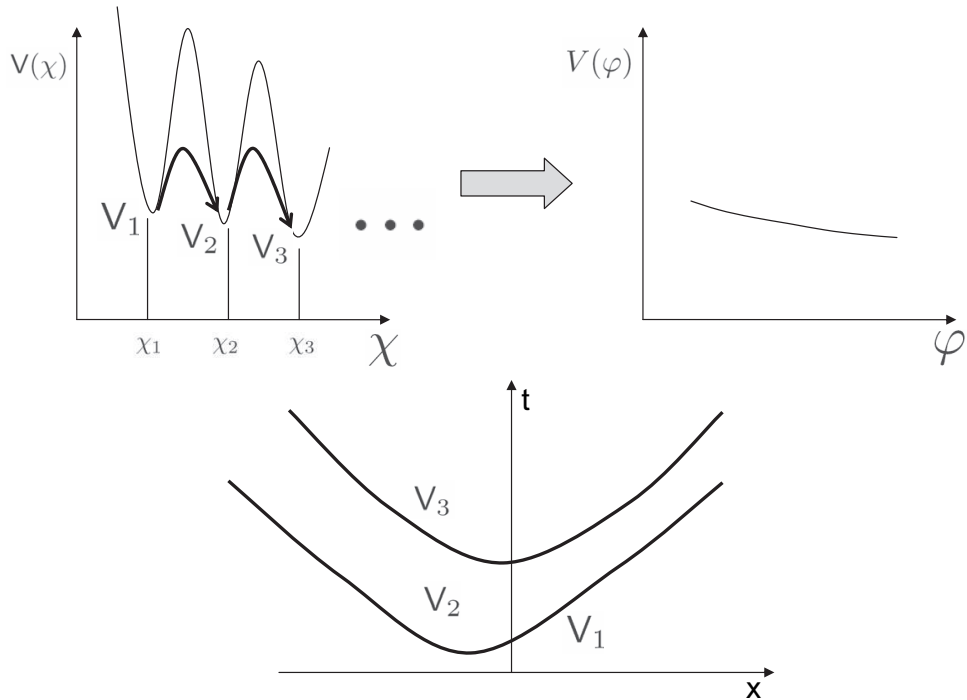}
\caption{\label{vv} {\it Top}: Correspondence between the potential
of an ordinary scalar field, ${\sf V}(\chi)$, and the effective {\it
Cuscuton} potential, $V(\v)$, which only passes through the minima
of ${\sf V}(\chi)$. {\it Bottom}: Corresponding space-time diagram
for consecutive tunneling events through minima of ${\sf V}(\chi)$}
\end{figure*}

In this section, we discuss whether {\it Cuscuton} action may be
derived from an ordinary field theory by integrating out degrees of
freedom.  In the process, we elucidate the physical intuition for this
theory.  We also discuss whether this type of action can be stable
against quantum corrections.

One may wonder whether one can introduce auxiliary fields that lead
to {\it Cuscuton} action after the constraints involving the
auxiliary fields are solved for.  Such systems can certainly be
written, but unless the auxiliary fields are made dynamical (by
giving them kinetic terms), such theories are identical to the
original theory.  We have tried several actions with auxiliary
fields whose solutions to the constraint equations leads to the {\it
Cuscuton} action.  Unfortunately, when dynamics (in the form of
canonical kinetic terms) are given to the auxiliary fields, even
after fine tuning of couplings and scales, the would-be auxiliary
fields settle to a field configuration different from the case when
the fields were non-dynamical.

Given that the action may have something to do with soap
films/bubbles, we have also tried to interpret the theory in terms
of integrating out short wavelength degrees of freedom which include
instanton transitions.  Imagine a regular field theory with a tilted
washboard potential that has an infinite number of discrete local
minima at $\chi_1, \chi_2, \chi_3, ...$, with the values ${\sf V}_1,
{\sf V}_2, {\sf V}_3, ...$, which monotonically decrease with the
field value. While classically, the field could have multiple vacua
at each minimum, none of them is stable under quantum tunneling.
Therefore, assuming a low tunneling probability, the full
evolutionary history of the field consists of a series of tunnelings
${\sf V}_1 \rightarrow {\sf V}_2 \rightarrow {\sf V}_3 \rightarrow
...$, , as shown in Fig. (\ref{vv}), plus small oscillations around
each minima following each tunneling event.  The Euclidian instanton
action for the successive tunnelings has the form \beq S_{\rm E} =
\int d^4x [\frac{1}{2}\partial_a\chi\partial_a\chi + {\sf V}(\chi)],
\eeq where $a=1..4$ counts over the 4-coordinates, including the
Euclidian time.

Now, focusing on long wavelength fluctuations with $k \ll (\Delta
V_{\rm max})^{1/2}/\Delta\chi$, which is equivalent to integrating out
modes that are shorter than the tunneling time, and in the thin-wall
approximation limit that the change in the potential ${\sf V}_i-{\sf
V}_{i+1}$ is much smaller than the height of the potential barrier,
$\Delta V_{\rm max}$ \cite{Coleman:1977th}, one may naively guess an
effective action of the form \beq S_{\rm E, eff} \simeq \sum_i J_i\int
d\Sigma_i + \int d^4x ~{\sf V}(\chi),
\label{eq:guessinstanton}
\eeq where \beq J_i \equiv
\int_{\chi_i}^{\chi_{i+1}} d\chi\sqrt{2[~{\sf V}(\chi)-{\sf V}_i]},
\eeq and $d\Sigma_i$ is the volume of the tunneling hypersurface for
the ${\sf V}_i$ to ${\sf V}_{i+1}$ transition
\cite{Coleman:1977py,Coleman:1977th}.

Now defining: \beq \mu^2(\v_{i+1}-\v_i) \equiv J_i =
\int_{\chi_i}^{\chi_{i+1}} d\chi\sqrt{2[~{\sf V}(\chi)-{\sf V}_i]},
\eeq and rotating back to the Minkowski coordinates, one would end up
with the action: \beq S_{\rm eff} \simeq \mu^2\sum_i (\v_{i+1}-\v_i)
\int d\Sigma_i - \int d^4x ~V(\v), \label{seff}\eeq where we also
define \beq V(\v_i)= {\sf V}(\chi_i). \eeq This is interesting since
Eq.~(\ref{seff}) becomes equivalent to the {\it Cuscuton} action
(Eq.~(\ref{sigma})) in the continuous limit.

Unfortunately, because Eq.~(\ref{eq:guessinstanton}) is valid only
for a somewhat ill-defined restricted class of field variations
close to those mimicking thin-walls, and because action is based on
analytic continuation into Euclidean space, the extent to which
Eq.~(\ref{eq:guessinstanton}) can be interpreted as that due to
integrating out tunneling transitions is unclear, at best.  Note
that the field path in the analytically continued saddle point
approximation does not have a simple connection with the space-time
picture of the process since the former is simply an approximation
scheme.  Hence, although tantalizing, we cannot make any rigorous
connection of {\it Cuscuton} to instanton induced bubble walls.

On the other hand, what we have learned from this exercise is that if
there is any quantum action which can be reduced to a set of discrete
degrees of freedom described by the action of the form
Eq.~(\ref{seff}), then this dynamics can be encoded by the {\it
Cuscuton} action in the limit that the number of discrete degrees of
freedom is large.

Furthermore, notice that the action Eq.~(\ref{sigma}) (or Eq.~(\ref{seff})) is
the most general local action that remains invariant under the field
transformations that preserve the area, $\Sigma(\v)$ and volume,
${\cal V}(\v)$, of constant field hypersurfaces, where \bea
\Sigma(\v_0) = \frac{\partial}{\partial\v_0}\int d^4x \sqrt{-g}
\sqrt{|\partial^\mu\v \partial_\mu\v|}~ \Theta(\v_0-\v), \\
{\cal V}(\v_0) = \frac{\partial}{\partial\v_0}\int d^4x \sqrt{-g}
~\Theta(\v_0-\v). \eea Therefore, it is reasonable to expect the
quantum corrections to be under control at low energies, where
higher order curvature effects could be neglected. This is also
consistent with the absence of quantum corrections to {\it Cuscuton}
action, as the linear perturbations have a degenerate phase space.

The attempted instanton picture for {\it Cuscuton} action involves a
potential, ${\sf V}(\chi)$, which is qualitatively very similar to
the recently proposed model known as Devaluation
\cite{Freese:2005pu} (or Chain inflation \cite{Freese:2004vs}). It
remains to be seen if their picture of rapid bubble nucleation has
any relevance to our picture of coherent {\it Cuscuton} evolution
\cite{companion}, in an appropriate limit, where the radiation
generated due to bubble collisions can be neglected.

\section{Conclusions} \label{conclude}

In this paper, we have analyzed the flat space field theoretic aspects
of {\it Cuscuton} action, which we define as a scalar field theory
whose kinetic term reduces to a total derivative in the homogenous
limit.  This theory was defined in this way and may be of interest to
cosmology because it is equivalent to a k-essence fluid with an
infinite speed of sound.

The surprise is that even though the dynamical equations seem to
admit superluminal signal propagation, there is \emph{no} physical
violation of causality, as perturbations have degenerate phase space
and thus transport no information.  That means this can be used to
modify gravity and other field theories to which it couples in a
novel manner since the fields have a kinetic term, unlike usual
Lagrange multipliers or auxiliary fields. We have also verified that
this modification does not lead to superluminal propagation in
ordinary scalar fields.

We have found some general interesting features to classical
solutions to this theory.  Hypersurfaces of constant field have
constant mean curvatures, which makes them the Minkowski space
analogs of soap films (or soap bubbles) in the Euclidian space.  We
can also solve the theory completely in 1+1 dimensions, and thereby
prove non-local degeneracy of phase space (which is equivalent to
the overconstraining behavior of Cauchy initial conditions), as well
as generic presence of singularities in the solutions. Extension of
these results to higher dimensions is a plausible conjecture.

As far as the qualitative behavior of the physics is concerned, {\it
Cuscuton} action can be viewed as a continuum limit of an action of
the form Eq.~(\ref{seff}) governing a large number of discrete degrees
of freedom.
%can be realized as the continuous long
%wavelength limit of consecutive instanton transitions (or tunneling
%events).
It is the most general (local) action that only depends on the area
and volume of constant field hypersurfaces, which is why it may
probably be protected against quantum corrections at low
energies. This is consistent with the absence of quantum corrections
to the {\it Cuscuton } action as a result of its degenerate phase
space.

Cosmology literature does not lack in abundance of models of dark
energy (or its alternatives).  The features that still make {\it
Cuscuton} interesting are the following:

\begin{itemize}
\item  Even though it has a kinetic term and modifies the cosmic dynamics, it does not introduce any additional (perturbative) degree of freedom.
Therefore, it can be considered a minimal modification of a
cosmological constant, or a minimal model for evolving dark energy.

\item It is protected against quantum corrections at low energies.
\end{itemize}
A companion paper \cite{companion} examines cosmology with a {\it
Cuscuton} dark energy fluid.

\begin{acknowledgments}
We would like to thank Nima Arkani-Hamed, Michael Doran, Gia Dvali,
and Aki Hashimoto for useful discussions. The work of DJHC and GG
was supported by DOE Outstanding Junior Investigator Program through
Grant No. DE-FG02-95ER40896 and NSF Grant No. PHY-0506002. NA wishes
to thank the hospitality of the Physics department at the University
of Wisconsin-Madison throughout the course of this project, as well
as the cosmocoffee.com weblog, where this project was originated.
\end{acknowledgments}

\bibliography{cuscutan7}

\appendix

\section{Propagating Waves in $X<0$ regime}\label{bubble}

In this section, we show that perturbations in an $X<0$ background
are similar to waves propagating on the surface of a bubble. To see
this, consider $\varphi_0(x)$ to be a stationary solution, thus
$u^i\equiv\nabla\varphi_0/|\nabla\varphi_0|$ represents the normal
vectors to two dimensional constant $\varphi_0$ CMC surfaces in
$3$-space.

Now perturbing the field around this solution
\begin{equation}
\varphi(x,t)=\varphi_0(x)+\delta\varphi(x,t), \end{equation} and
substituting $\varphi(x,t)$ back into the equation of motion
(\ref{field}), in the short wavelength limit where the changes in
$|\nabla\varphi_0|$ can be neglected, we find:
\begin{equation}
\partial_\mu\partial^\mu
\delta\varphi(x,t)+u_iu_j\partial^i\partial^j\delta\varphi(x,t)+
\frac{V''(\varphi_0)}{\mu^{2}}|\nabla\varphi_0|\delta\varphi(x,t)\simeq0.\end{equation}
Note that $\mu^2<0$ to ensure energy positivity, as discussed in
Sec. (\ref{causality}). In Fourier space $\delta\varphi(x,t)=\int
{d^4k\over(2\pi)^4}\varphi_k e^{k_\mu x^\mu}$, this equation imposes
the following dispersion relation modes
\begin{equation}\omega^2-|{\bf k}_{\|}|^2+\frac{V''(\varphi_0)}{\mu^{2}}|\nabla\varphi_0|=0,\label{disp}\end{equation}
where $\omega=k^0$ and ${\bf k}_\| $ is the component of spatial
wavevector parallel to constant field surface. Eq. (\ref{disp})
describes modes that propagate along the surfaces of constant $\v$
and are thus analogous to waves on the surface of a bubble.

\end{document}